\documentclass[preprint,authoryear]{elsarticle}

\usepackage{amssymb}
\relax
\def\frac#1#2{{#1\over#2}}
\def\<{\langle}\def\>{\rangle}
\newcount\parenthesis \parenthesis=0 \newcount\n
\def\({\global\advance\parenthesis by1\left(}
\def\){\global\advance\parenthesis by-1\right)}
\def\{{\global\advance\parenthesis by1\left\lbrace}
\def\}{\global\advance\parenthesis by-1\right\rbrace}
\def\[{\relax} % dummy parenthesis
\def\]{\relax} % dummy parenthesis
\def\Loop#1\Repeat{\global\n=0\global\let\body=#1\iterate}
\def\iterate{\body\let\next=\iterate\else\let\next=\relax\fi\next}
\def\ldd{\ifnum\n<\parenthesis\global\advance\n by1
\left.\nulldelimiterspace=0pt\mathsurround=0pt}
\def\rdd{\ifnum\n<\parenthesis\global\advance\n by1
\right.\nulldelimiterspace=0pt\mathsurround=0pt}
\def\nl{\Loop\rdd\Repeat\hfill\cr\qdd\Loop\ldd\Repeat{}}

\def\qdd{\quad\quad}
\newcount\exacount\exacount=0\font\caps=cmcsc10
\def\Istrut{\vrule height11pt depth4pt width0pt}
\def\TRIexa#1#2#3#4{\global\advance\exacount by1\par\filbreak
{\offinterlineskip
  \vbox{\hrule\hbox to\hsize{\Istrut\vrule
      \hbox to 8mm{\hfil\caps\the\exacount\hfil}\vrule
      \quad\rm#1\hfill\vrule
      \hbox to 32mm{\hfill{\caps Mode: }{\tt #2}\hfill}\vrule
      \hbox to 32mm{\hfill{\caps Tolerance: }{\tt #3}\hfill}\vrule}
    \hrule\hbox to\hsize{\Istrut\vrule\hfill#4\hfill\vrule}\hrule}
}\nobreak}
\renewcommand{\cdot}{\,}

\journal{Computational Statistics and Data Analysis}

\begin{document}

\newtheorem{thm}{Theorem}
\newtheorem{lem}[thm]{Lemma}
\newtheorem{defn}{Definition}
\newdefinition{rmk}{Remark}
\newproof{pf}{Proof}
\newproof{pot}{Proof of Theorem \ref{thm2}}

\begin{frontmatter}



%\title{A measure of skewness of testing for normality}
\title{A measure of skewness for testing departures from normality}


\author[NAKAGAWA]{Shigekazu Nakagawa\corref{cor1}}
\cortext[cor1]{Corresponding author.}
\ead{nakagawa@cs.kusa.ac.jp}
\address[NAKAGAWA]{
Department of Computer Science and Mathematics, 
Kurashiki University of Science and the Arts,
2640 Nishinoura, Tsurajima-cho, Kurashiki-shi,
Okayama 712-8505, JAPAN}
% \author[KURODA]{Masahiro Kuroda\corref{cor1}}
% \cortext[cor1]{Corresponding author.}
% \ead{kuroda@soci.ous.ac.jp}
% \address[KURODA]
% {Department of Socio-Information, Okayama University of Science,
% 1-1 Ridaicho, Kita-ku, Okayama 700-0005, JAPAN}
%
\author[HASHIGUCHI]{Hiroki Hashiguchi}
\address[HASHIGUCHI]
{Department of Information and Computer Sciences, Saitama University,
255 Shimo-Okubo, Sakura,
Saitama  338-8570, JAPAN}
%
% \author[NAKAGAWA]{Shigekazu Nakagawa}
% \address[NAKAGAWA]{
% Department of Computer Science and Mathematics, 
% Kurashiki University of Science and the Arts,
% 2640 Nishinoura, Tsurajima-cho, Kurashiki-shi,
% Okayama 712-8505, JAPAN}
%
% \author[GENG]{Zhi Geng}
% \address[GENG]{
% Institute of Mathematical Sciences,
% Peking University, Beijing 100871, CHINA}
\author[NIKI]{Naoto Niki}
\address[NIKI]
{Department of Management Science, 
Tokyo University of Science, 
1-3 Kagurazaka, Shinjuku-ku, Tokyo 162-8601, JAPAN}

\begin{abstract}
We propose a new skewness test statistic 
for normality based on the Pearson measure of skewness. 
We obtain asymptotic first four moments of the null distribution for this statistic  
by using a computer algebra system and 
its normalizing transformation based on the Johnson $S_{U}$ system.
Finally the performance of the proposed statistic is shown  
by comparing the powers of several skewness test statistics against some alternative hypotheses.
\end{abstract}

\end{frontmatter}

%\linenumbers

\section{Introduction}
A test for normality is an essential problem in statistical practice.
Earlier studies on tests for normality are summarized in \cite{Thode}.
Traditionally, it is common to use skewness and kurtosis statistics $\surd{b_1}=m_{3}/m_{2}^{3/2}$ 
and $b_2 =m_{4}/m_{2}^{2}$, respectively, 
where for a random sample $(X_{1}, X_{2}, \ldots , X_{n})$, 
$m_{r} = (1/n)\sum_{i=1}^{n}\left(X_{i} - \bar{X} \right)^{r}$, $r=2, 3$, and $4$, 
and $\bar{X} = (1/n)\sum_{i=1}^{n}X_{i}$. 
They are used for detecting skew symmetric and heavy or light tails, respectively.  
The Jarque--Bera test combining $\surd{b_1}$ and ${b_2}$ 
is well known as an omnibus test for normality (\cite{MR963337}). 
The improved omnibus test was recently presented in \cite{NHN}.
The above test statistics are all based on sample moments.
Furthermore, the Shapiro--Wilk ($W$) test is famous 
as an extension of probability plots (\cite{shapiro1965analysis}).
 
In addition, the Lin--Mudholkar ($LM$) test is developed for asymmetric alternatives (\cite{LinMudholker}). 
The focus of this paper is also on tests that detect skew symmetric. 
Based on $\surd{b_1}$ and ${b_2}$, 
we propose a test statistic 
\begin{equation}
spms = \frac{\surd{b_{1}}\left( b_2 + 3\right)}{2\left( 5b_{2}- 6b_{1}-9\right)},
\label{spms}
\end{equation}
and called the sample Pearson measure of skewness ($spms$).
The statistic $spms$ corresponds to the Pearson measure of skewness defined by 
\begin{equation}
pms = 
%(\frac{\mbox{mean} - \mbox{mode}}{\sigma}
%=) 
\frac{\surd{\beta_{1}}\left( \beta_2 + 3\right)}%
{2\left( 5\beta_{2}-6\beta_{1}-9\right)}, 
\label{pms}
\end{equation}
where $\surd{\beta_{1}} = \mu_{3}/\mu_{2}^{3/2}$ and $\beta_{2}=\mu_{4}/\mu_{2}^{2}$,  
and for $r=2,3,4$, $\mu_{r}$ denotes the $r$th population moment about the mean.
The $pms$ is introduced by Karl Pearson considering Pearson Systems which 
are distinguished nine types including normal distributions (\cite{SO}). 
Under Pearson Systems, the $pms$ actually coincides with a measure of skewness
$$\frac{\mbox{mean} - \mbox{mode}}{\sigma},$$
where $\sigma$ is the standard deviation.
As the mean and the mode coincide in a 
symmetric population, the distance from the mean to the mode can be treated 
as a measure of skewness.

The first four moments of the null distribution for $spms$ are explained in Section~\ref{pr}.
Based on the Johnson $S_{U}$ system, 
we obtain a transformation to approximate the normality of 
the null distribution of $spms$ in Section~\ref{tr}.
In Section~\ref{num_il}, the performance of $spms$ is shown  
by comparing the powers of several skewness test statistics 
against six asymmetric alternatives.

\section{Approximate moments of the null distribution for $spms$}
\label{pr}

\begin{figure}[t]
\centerline{
\includegraphics[width=0.805\textwidth]{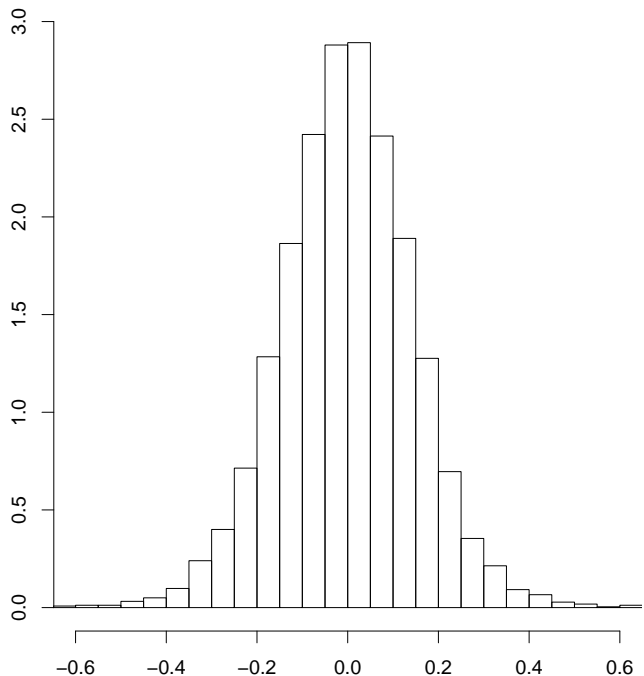}
}
% ./Rprogs/drawhist.txt
\caption{Histogram of $spms$ with $n=100$ ($10^4$ replications)}
\label{hist_spms}
\end{figure}

In this section, we assume that 
a sample $(X_{1}, X_{2}, \ldots , X_{n})$ is drawn from a normal population.
The $spms$ is invariant under the origin, and the scale changes 
because of the invariance of $\surd{b_{1}}$ and $b_{2}$. 
It is clear that $spms$ is symmetric. 
Figure~\ref{hist_spms} shows a histogram of $spms$ with $n=100$.

To find the approximate moments of the distribution for $spms$, 
a large amount of symbolic computation is required.
\cite{NN} described a method for this purpose as follows:
\begin{itemize}
\item[(1)]
Set
$
U = \sqrt{n}\left( m_{2} - 1 \right), \,
V = \sqrt{n}m_{3}, \,
W = \sqrt{n}\left( m_{4} - 3 \right),
$
and expand $spms$ as power series in terms of $1/\sqrt{n}$:
$$\displaylines{
spms = \frac{1}{\sqrt{n}}
\(
  \frac{1}{
        2}\cdot V
        \)
  +\frac{1}{n}\cdot 
  \(\frac{5}{
          4}\cdot U\cdot V
    -
    \frac{1}{
          3}\cdot V\cdot W
  \)
  +\frac{1}{n\sqrt{n}}\cdot 
  \(\frac{79}{
          16}\cdot U^{2}\cdot V
\nl\hspace*{-3em}
    -
    \frac{13}{
          6}\cdot U\cdot V\cdot W
    +
    \frac{1}{
          2}\cdot V^{3}
    +
    \frac{5}{
          18}\cdot V\cdot W^{2}
  \)
%\nl\hspace*{-3em}
+
\frac{1}{n^2}\cdot 
  \(\frac{517}{
          32}\cdot U^{3}\cdot V
    -
    \frac{263}{
          24}\cdot U^{2}\cdot V\cdot W
\nl\hspace*{-3em}
    +
    \frac{9}{
          4}\cdot U\cdot V^{3}
    +
    \frac{95}{
          36}\cdot U\cdot V\cdot W^{2}
    -
    \frac{3}{
          4}\cdot V^{3}\cdot W
    -
    \frac{25}{
          108}\cdot V\cdot W^{3}
  \)
 + O\(n^{-5/2} \).
}$$
We remark that $U = O(1)$, $V = O(1)$, and $W = O(1)$.
\item[(2)]
Take expectations term by term.
Thus, we obtain the first approximate moment of $spms$.
\item[(3)]
The second, third, and fourth moments are obtained 
from  $spms^{2}$,  $spms^{3}$, and $spms^{4}$, respectively.
\end{itemize}
In step (2), each $U^{i}V^{j}W^{k} \, (0 \leq i+j+k \leq 8)$ 
is a symmetric polynomial of $X_{1}, X_{2}, \ldots , X_{n}$.
We emphasize that 
the algorithm for the change of bases of symmetric polynomials 
described in \cite{NN} is applied in computing 
expectations $\mbox{E}\left[U^{i}V^{j}W^{k}\right]$.
The algorithm has been implemented in REDUCE (\cite{H}) with utility programs.

Let $\lambda_{r}$ be the $r$th ($r=2, 4$) moment of the distribution for $spms$ 
drawn from a normal population, each having the following form:
\begin{eqnarray}
% \lambda_{1} &=& 0, \qquad
\lambda_{2} &=& \left(\frac{3}{2}\right)n^{-1} 
  + 41n^{-2} + \left(\frac{6511}{2}\right)n^{-3} + O\left( n^{-4}\right),
\label{var}
\\
% \lambda_{3} &=& 0, \qquad
\lambda_{4} &=& \left(\frac{27}{4}\right)n^{-2} +414n^{-3} + O\left( n^{-4}\right).
\label{lambda4}
\end{eqnarray}
Note that all odd moments are zero.
From (\ref{var}), $spms$ is asymptotically normally distributed with mean 0 and 
variance $3/(2n)$.
Thus, we obtain 
\begin{equation}
\beta_{2}(spms) = \frac{\lambda_{4}}{\lambda_{2}^{2}} 
= 3 + 20n^{-1} + \left(\frac{48544}{3}\right)n^{-2} 
+ \left(\frac{10386704}{9}\right)n^{-3} + O\left( n^{-4}\right),  
\end{equation}
which will be used for obtaining a normalizing transformation.

\section{Transformation}
\label{tr}

A normal approximation of the null distribution for $spms$ is obtained
using a transformation described in \cite{J}, given as follows:
\begin{thm}
Assume that the null hypothesis of normality is true 
and let
\begin{equation}
Y = \frac{spms}{\sqrt{\lambda_{2}}},
\label{y:ex}
\end{equation}
\begin{equation}
W^{2} = -1 + \sqrt{2\left( \beta_{2}(spms) - 1\right)},
\end{equation}
\begin{equation}
\delta = \frac{1}{\sqrt{\log W}},
\end{equation}
\begin{equation}
\alpha = \sqrt{\frac{1}{W^2 - 1}}.
\label{alpha:ex}
\end{equation}
Then
\begin{equation}
Z =\delta\log\left( \frac{Y}{\alpha} + 
\sqrt{ 1 + \left( \frac{Y}{\alpha}\right)^{2}} \right)
\label{Z:ex}
\end{equation}
is approximately a standard normal variate with mean zero and variance unity.
\end{thm}
This transformation is an analogy of the normalizing transformation of the 
null distribution for $\surd{b_{1}}$ given by \cite{DA}. 

\begin{figure}[t]
\centerline{
\includegraphics[width=0.805\textwidth]{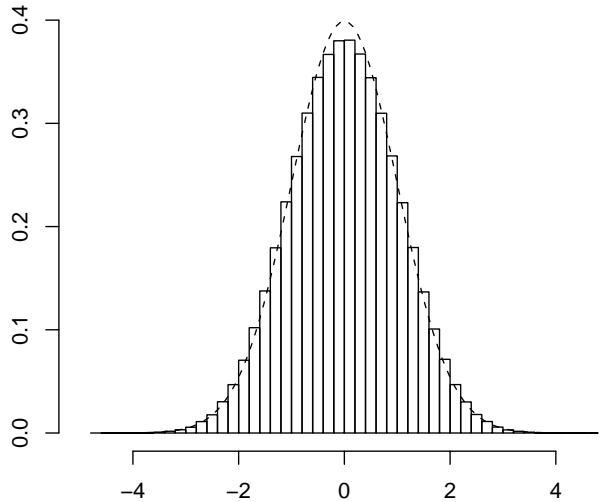}
}
\caption{Histogram of $Z$ (\ref{Z:ex}) with $n=200$ ($10^6$ replications) 
and the standard normal probability density function (dashed curve)}
\label{pdf}
\end{figure}

Figure~\ref{pdf} shows a histogram of $Z$ with $n=200$ 
when $10^6$ random normal samples are generated.  
The dashed line denotes the probability density function of the standard normal distribution.
Both one--sided and two--sided tests at any desired levels of significance can be performed.
For example, for a two--sided test with a $0.05$ level of 
significance, the null hypothesis is rejected if $\left| Z \right| > 1.96$.

% \section{Accuracy of the transformation}
% \label{ac}

Monte Carlo simulations are performed for judging the accuracy of the transformation.
For samples of sizes $n=100, 150, 200, 300, 500$, and $1000$, 
$10^6$ random normal samples are generated.
For each sample, the $spms$ statistic is computed and classified into the intervals 
formed by the two-sided tests with levels of significance $0.01$, $0.05$, $0.10$, and $0.20$.
The entries in Table~\ref{tab:MC} show  
the Monte Carlo relative frequencies for the intervals.
For $n \geq 500$, the Monte Carlo results agree to at least two decimal places for any significance level.
Even if $n = 100$, the results are good for $0.01$ level of significance.

\begin{table}[pb]
\begin{center}
\caption{Monte Carlo probabilities for various levels of significance using $Z$ (\ref{Z:ex})
with $10^6$ replications per sample size}
\label{tab:MC}
\vspace*{1em}
\begin{tabular}{crrrrrr} 
Presumed levels & \multicolumn{6}{c}{sample size} \\
two-sided test & 100 & 150 & 200 & 300 & 500 & 1000 \\
\hline
0.01 & 0.0112  & 0.0099  & 0.0098  & 0.0098  & 0.0102  & 0.0100  \\
0.05 & 0.0595  & 0.0537  & 0.0519  & 0.0513  & 0.0504  & 0.0501  \\
0.10 & 0.1207  & 0.1097  & 0.1057  & 0.1027  & 0.1009  & 0.1003  \\
0.20 & 0.2368  & 0.2191  & 0.2113  & 0.2052  & 0.2019  & 0.2004  \\
\end{tabular}
\end{center}
\end{table}

\section{Power study}
\label{num_il}

This section demonstrates a comparison of the powers of $spms$ 
with $\surd{b_{1}}$, $W$, and $LM$ tests.
Monte Carlo simulations 
are conducted 
with samples of sizes $n = 40, 50, 60, 80$, and $100$ and $10^4$ replications.
Two--sided tests with a $5\%$ significance level are carried out 
against six alternative distributions:
beta distributions with two parameters $(p, q) = (2, 1), (3, 2)$;
gamma distributions with shape parameters $\alpha = 2, 3$;
Weibull distributions with shape parameters $\alpha = 2$;
and log--normal distribution with mean $\mu=0$ and standard deviation $\sigma=1/2$.
The first two distributions have light tails and the rest have heavy tails.

\begin{figure}[htbp]
\centerline{
\includegraphics[width=0.805\textwidth]{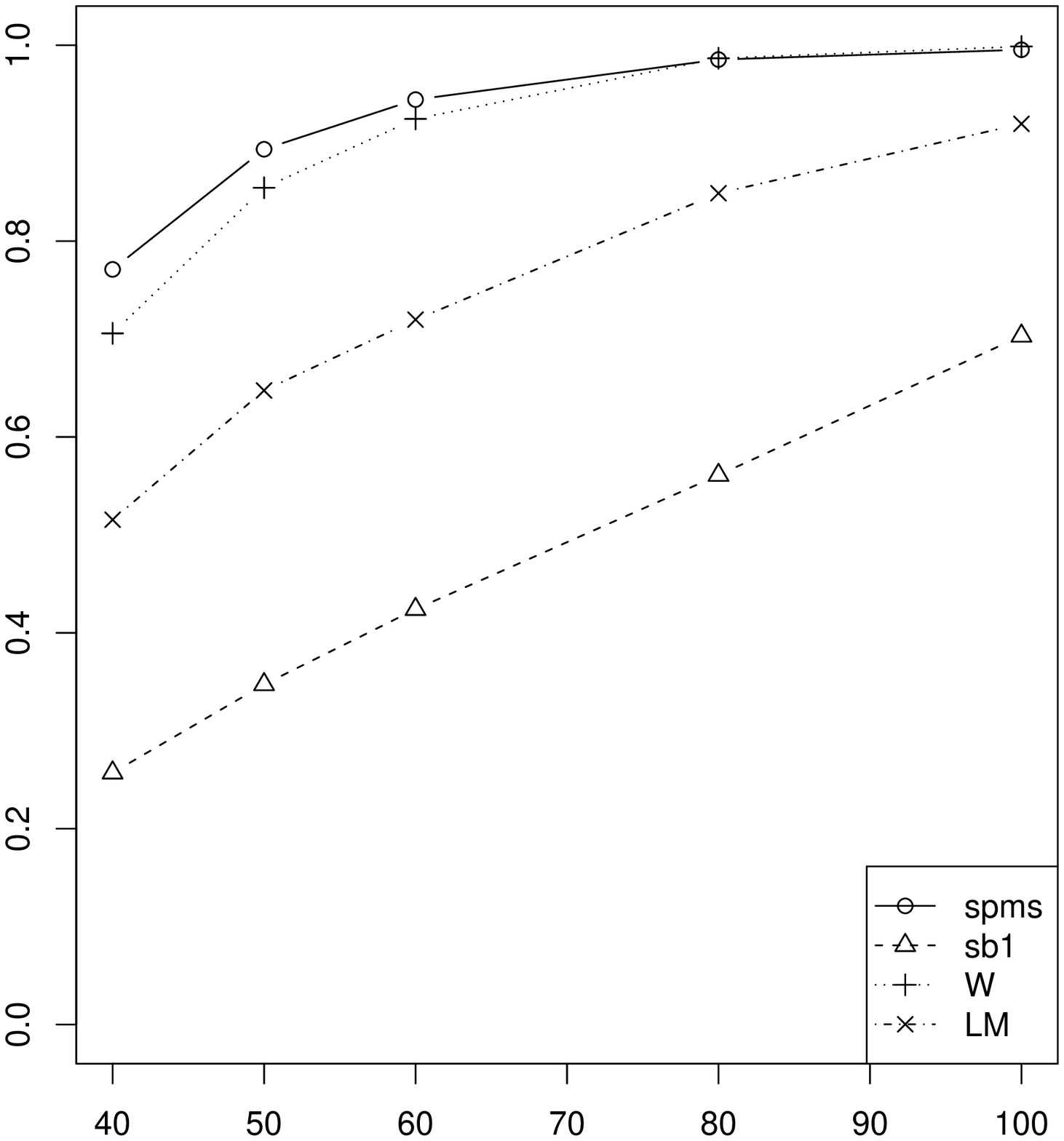}
}
\caption{Powers of the tests when the alternative distribution is Beta $(\alpha=2, \beta=1)$}
\label{pc:beta21}
\end{figure}

\begin{figure}[htbp]
\centerline{
\includegraphics[width=0.805\textwidth]{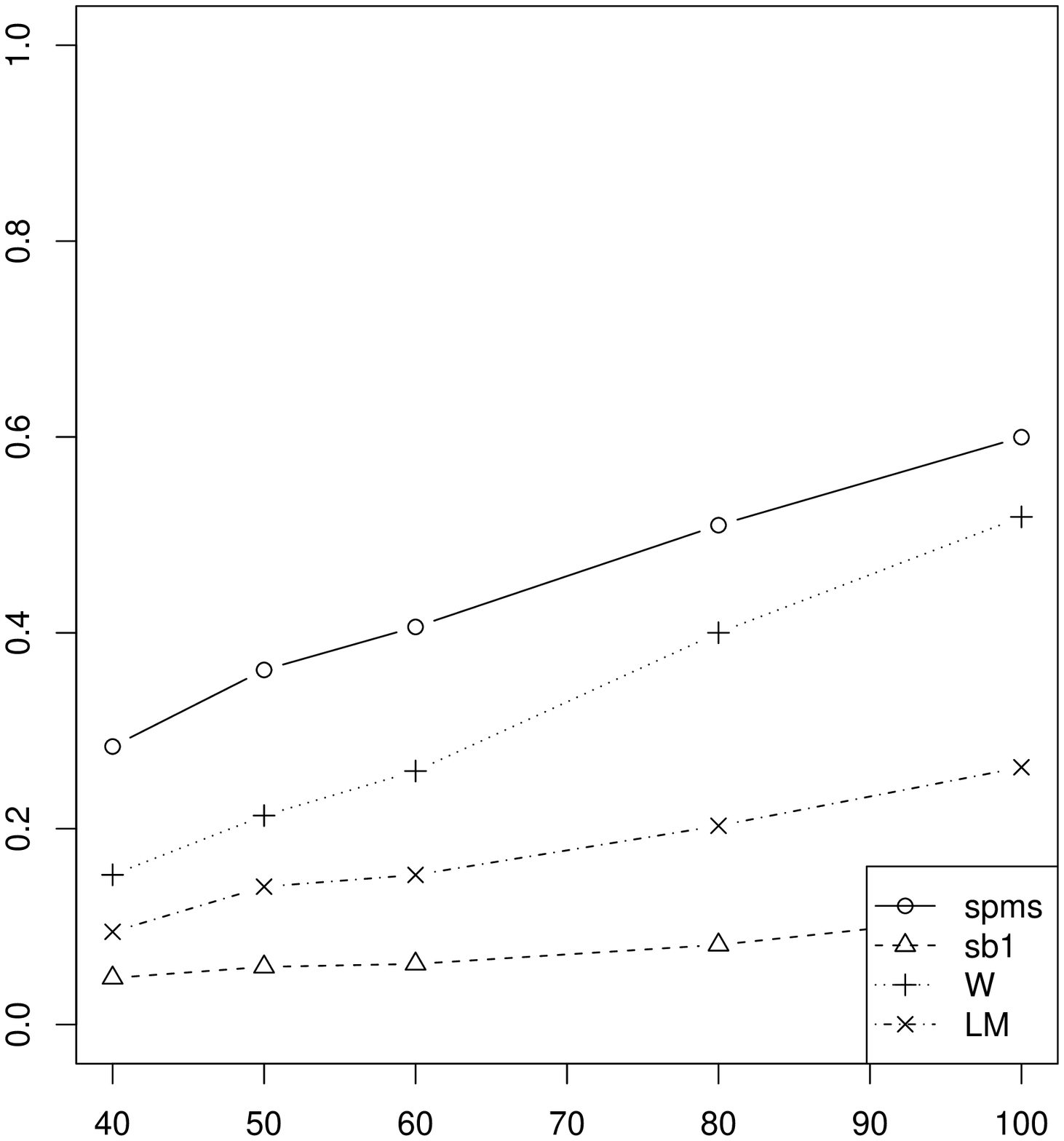}
}
\caption{Powers of the tests when the alternative distribution is Beta $(\alpha=3, \beta=2)$}
\label{pc:beta32}
\end{figure}

\begin{figure}[htbp]
\centerline{
\includegraphics[width=0.805\textwidth]{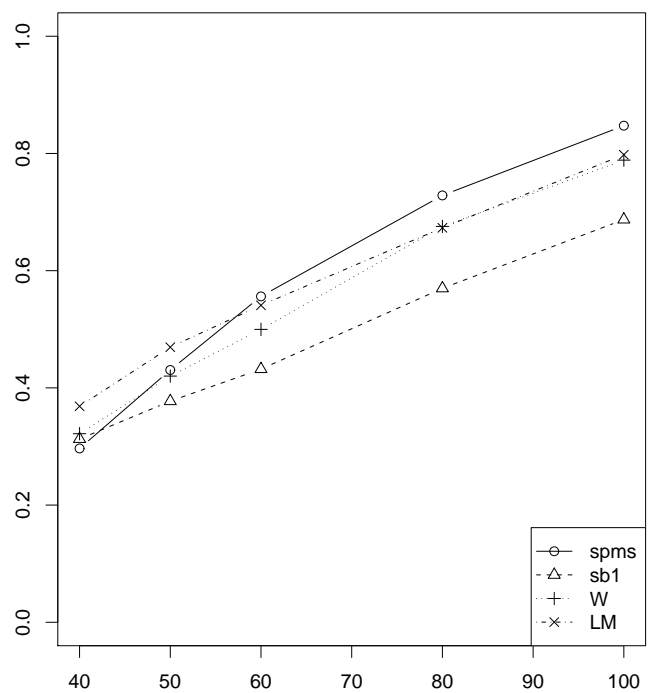}
}
\caption{Powers of the tests when the alternative distribution is Weibull $(\alpha=2)$}
\label{pc:weibull2}
\end{figure}

\begin{figure}[htbp]
\centerline{
\includegraphics[width=0.805\textwidth]{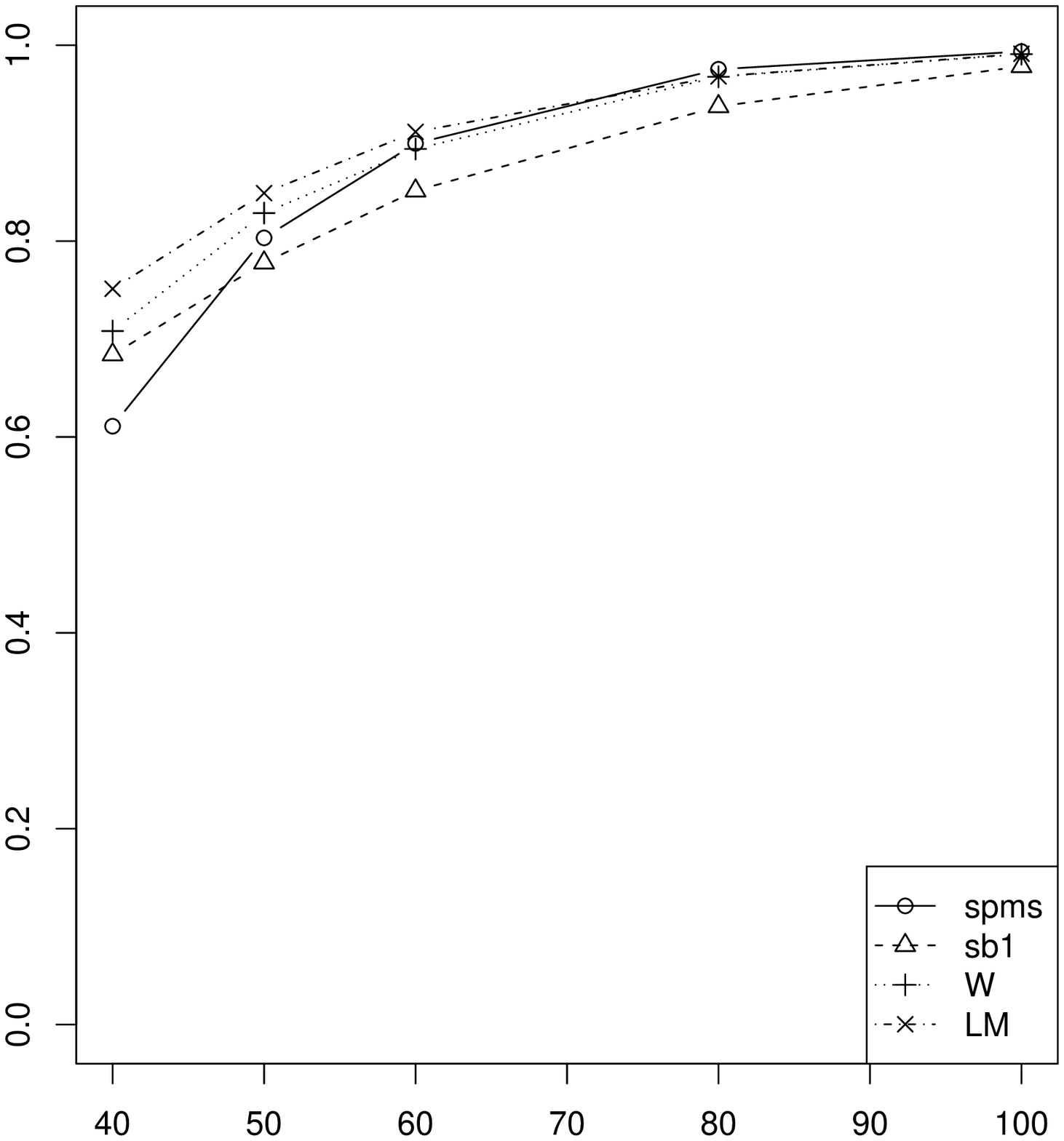}
}
\caption{Powers of the tests when the alternative distribution is Gamma $(\alpha=3)$}
\label{pc:gamma3}
\end{figure}

\begin{figure}[htbp]
\centerline{
\includegraphics[width=0.805\textwidth]{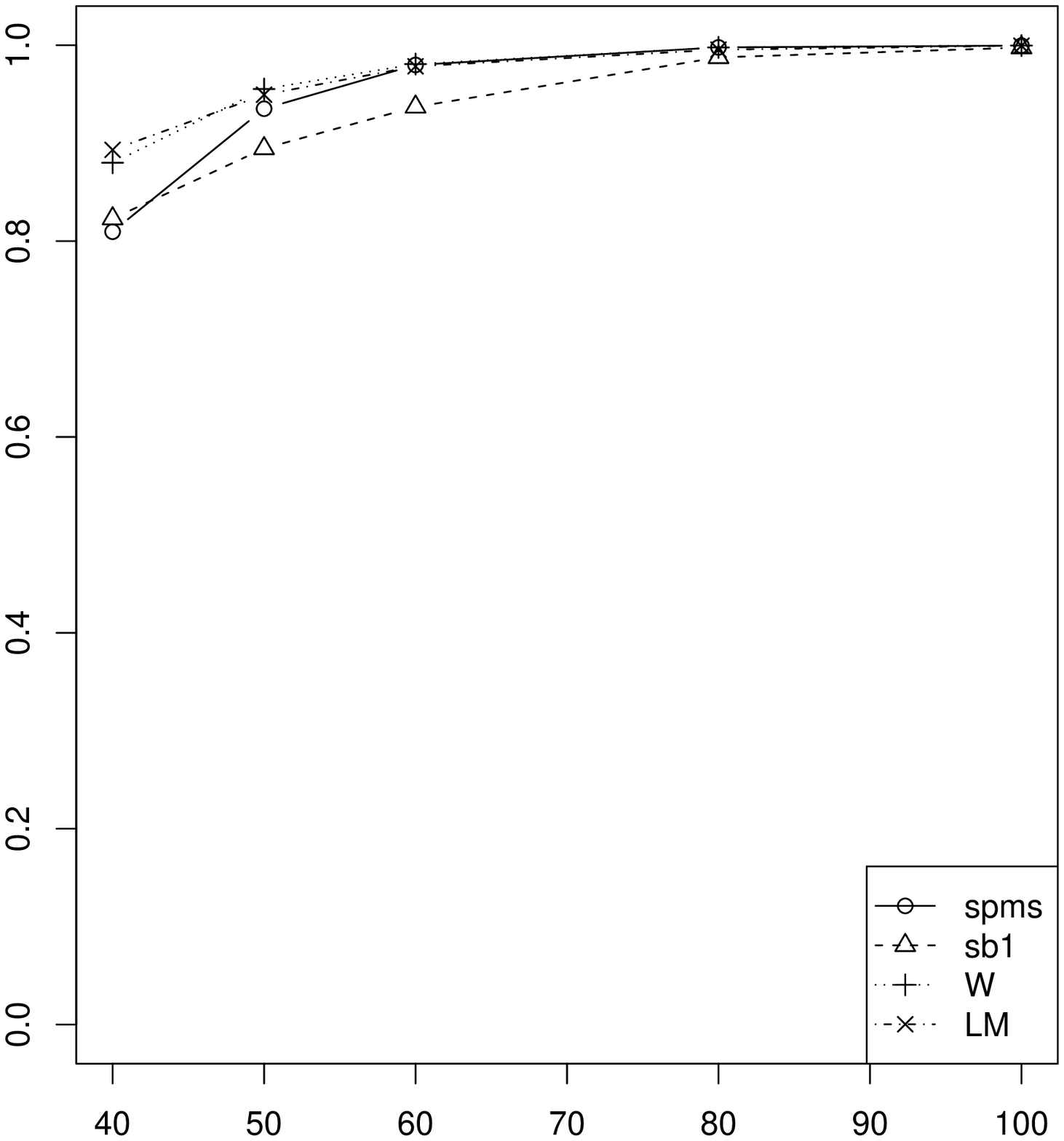}
}
\caption{Powers of the tests when the alternative distribution is Gamma $(\alpha=2)$}
\label{pc:gamma2}
\end{figure}

\begin{figure}[htbp]
\centerline{
\includegraphics[width=0.805\textwidth]{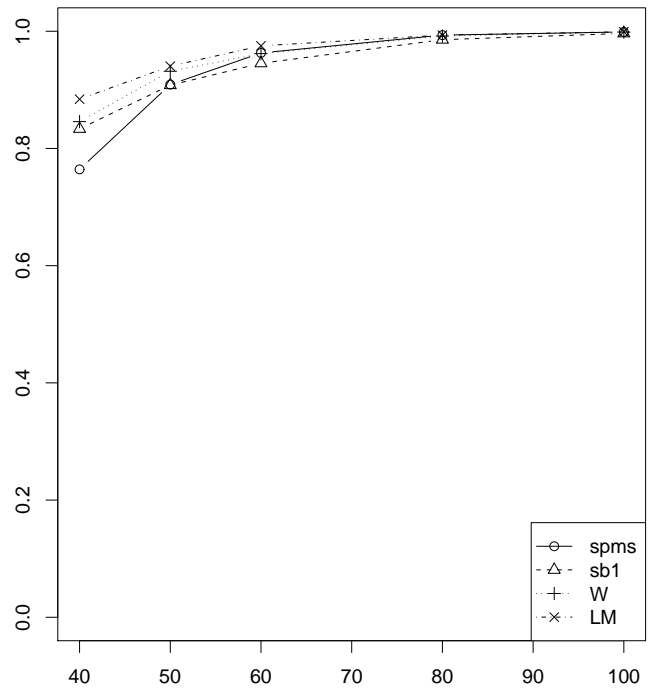}
}
\caption{Powers of the tests when the alternative distribution 
is log normal $(\mu=0, \sigma=1/2)$}
\label{pc:ln0half}
\end{figure}

The powers of the above four statistics are 
summarized in Table~\ref{tab:powers} and 
Figures~\ref{pc:beta21}--\ref{pc:ln0half} 
($spms$($\circ$), $\surd{b_{1}}$($\triangle$), $W$($+$), and $LM$($\times$)). 
In most cases, the power of the $spms$ test is better than that of the $\surd{b_{1}}$ test.
The power of the $spms$ test is superior to that of all other tests 
in the first two cases.
The $spms$ test is comparable when the alternative distributions have heavy tails. 

\begin{table}
\caption{Monte Carlo powers of some tests for normality, level of significance $0.05$, population skewness $\surd{\beta_{1}}$, and kurtosis $\beta_{2}$}
\label{tab:powers}
\begin{center}
\begin{tabular}{lrrrrrrrr}
  \hline
Alternative hypotheses & $\surd{\beta_{1}}$ & $\beta_{2}$ & $n$ & $spms$ & $\surd{b_1}$ & $SW$ & $LM$ \\ 
  \hline
Beta $(\alpha=2, \beta=1)$ & $-0.57$ & 2.4 & 40 & 0.771 & 0.257 & 0.706 & 0.515 \\ 
				&&	& 50 & 0.894 & 0.347 & 0.855 & 0.647 \\ 
				&&	& 60 & 0.945 & 0.424 & 0.925 & 0.720 \\ 
				&&	& 80 & 0.985 & 0.561 & 0.987 & 0.849 \\ 
				&&	& 100 & 0.995 & 0.703 & 0.999 & 0.920 \\ 
Beta $(\alpha=3, \beta=2)$ & -0.29 & 2.4  & 40 & 0.284 & 0.048 & 0.153 & 0.095 \\ 
				&&	& 50 & 0.362 & 0.059 & 0.213 & 0.141 \\ 
				&&	& 60 & 0.406 & 0.062 & 0.259 & 0.153 \\ 
				&&	& 80 & 0.510 & 0.081 & 0.400 & 0.203 \\ 
				&&	& 100 & 0.600 & 0.114 & 0.518 & 0.263 \\ 
Wiebull $(\alpha=2)$		& 0.63  & 3.3 & 40 & 0.297 & 0.313 & 0.322 & 0.369 \\ 
				&&	& 50 & 0.430 & 0.377 & 0.420 & 0.469 \\ 
				&&	& 60 & 0.556 & 0.432 & 0.500 & 0.541 \\ 
				&&	& 80 & 0.728 & 0.570 & 0.675 & 0.673 \\ 
				&&	& 100 & 0.847 & 0.687 & 0.789 & 0.798 \\ 
Gamma $(\alpha=3)$		& 1.16  & 5.0 & 40 & 0.611 & 0.684 & 0.708 & 0.751 \\ 
				&&	& 50 & 0.803 & 0.778 & 0.829 & 0.849 \\ 
				&&	& 60 & 0.900 & 0.851 & 0.894 & 0.912 \\ 
				&&	& 80 & 0.976 & 0.938 & 0.968 & 0.968 \\ 
				&&	& 100 & 0.994 & 0.978 & 0.991 & 0.991 \\ 
Gamma $(\alpha=2)$		& 1.41  & 6.0 & 40 & 0.810 & 0.823 & 0.880 & 0.893 \\ 
				&&	& 50 & 0.935 & 0.895 & 0.955 & 0.950 \\ 
				&&	& 60 & 0.980 & 0.937 & 0.981 & 0.979 \\ 
				&&	& 80 & 0.998 & 0.988 & 0.998 & 0.996 \\ 
				&&	& 100 & 1.000 & 0.998 & 1.000 & 1.000 \\ 
%   26 & 40 & 0.978 & 0.963 & 0.995 & 0.991 \\ 
%   27 & 50 & 0.998 & 0.989 & 1.000 & 0.998 \\ 
%   28 & 60 & 1.000 & 0.997 & 1.000 & 1.000 \\ 
%   29 & 80 & 1.000 & 1.000 & 1.000 & 1.000 \\ 
%   30 & 100 & 1.000 & 1.000 & 1.000 & 1.000 \\ 
%   31 & 40 & 0.978 & 0.968 & 0.996 & 0.990 \\ 
%   32 & 50 & 0.998 & 0.989 & 1.000 & 0.998 \\ 
%   33 & 60 & 1.000 & 0.996 & 1.000 & 1.000 \\ 
%   34 & 80 & 1.000 & 0.999 & 1.000 & 1.000 \\ 
%   35 & 100 & 1.000 & 1.000 & 1.000 & 1.000 \\ 
%   36 & 40 & 0.431 & 0.732 & 0.824 & 0.638 \\ 
%   37 & 50 & 0.507 & 0.776 & 0.898 & 0.692 \\ 
%   38 & 60 & 0.558 & 0.807 & 0.936 & 0.716 \\ 
%   39 & 80 & 0.632 & 0.850 & 0.976 & 0.757 \\ 
%   40 & 100 & 0.682 & 0.873 & 0.990 & 0.798 \\ 
%   41 & 40 & 0.142 & 0.433 & 0.440 & 0.350 \\ 
%   42 & 50 & 0.195 & 0.468 & 0.526 & 0.386 \\ 
%   43 & 60 & 0.236 & 0.503 & 0.583 & 0.425 \\ 
%   44 & 80 & 0.286 & 0.566 & 0.691 & 0.460 \\ 
%   45 & 100 & 0.335 & 0.613 & 0.757 & 0.514 \\ 
log normal $(\mu=0, \sigma=1/2)$	& 1.80  & 8.9 	& 40 & 0.764 & 0.833 & 0.846 & 0.884 \\ 
				&	&      	& 50 & 0.909 & 0.908 & 0.932 & 0.940 \\ 
				&	&      	& 60 & 0.963 & 0.945 & 0.962 & 0.975 \\ 
				&	&	& 80 & 0.993 & 0.986 & 0.992 & 0.993 \\ 
				&	&	& 100 & 0.999 & 0.996 & 0.999 & 0.999 \\ 
   \hline
\end{tabular}
\end{center}
\end{table}
% ./Rprogs/mktabS3.r		...パーセント点と検出力計算プログラム
% ./Rprogs/Res_4/a05.txt	...検出力結果，有意水準5%
% ./pt_cal_spms_4.txt		...spmsパーセント点，10^4回
% ./pt_cal_sb1_4.txt		...sb1パーセント点，10^4回
% ./pt_cal_SW_4.txt			...SWパーセント点，10^4回
% ./pt_cal_Z_4.txt          ...Zパーセント点，10^4回

\section{Concluding remarks}
Corresponding to the Pearson measure of skewness $pms$ (\ref{pms}), we propose 
$spms$ (\ref{spms}) as a skewness test statistic for normality. 
We obtain the normalizing transformation (\ref{Z:ex}) of the null distribution for $spms$ 
based on the second and fourth approximate moments shown in (\ref{var}) and (\ref{lambda4}).
For a moderate sample size, such as $n \geq 200$, the transformation is 
valid in practice, as shown in Table~\ref{tab:MC}.

\section*{Acknowledgement}
This research is partially supported by the Japan Society for the 
Promotion of Science (JSPS), Grant--in--Aid for Scientific Research (C), 
No 22500266.

\end{document}

